\documentclass[12pt]{article}

\usepackage{graphics}
\usepackage{epsfig}
\usepackage{epsf,afterpage}
\usepackage{amssymb}
\newcommand{\beq}{\begin{eqnarray}}
 \newcommand{\eeq}{\end{eqnarray}}
\newcommand{\be}{\begin{equation}}
 \newcommand{\ee}{\end{equation}}
 \def\la{\mathrel{\mathpalette\fun <}}
\def\ga{\mathrel{\mathpalette\fun >}}
\def\fun#1#2{\lower3.6pt\vbox{\baselineskip0pt\lineskip.9pt
\ialign{$\mathsurround=0pt#1\hfil ##\hfil$\crcr#2\crcr\sim\crcr}}}

\newcommand{{\SD}}{\rm SD}

\newcommand{\vesig}{\mbox{\boldmath${\rm \sigma}$}}

\newcommand{\vez}{\mbox{\boldmath${\rm z}$}}

\newcommand{\vew}{\mbox{\boldmath${\rm w}$}}

\newcommand{\veB}{\mbox{\boldmath${\rm B}$}}

\newcommand{\veE}{\mbox{\boldmath${\rm E}$}}

\newcommand{\lan}{\langle}
\newcommand{\ran}{\rangle}

\begin{document}




\begin{titlepage}

\begin{center}

{\Large \bf Confinement Mechanism }

\medskip

{\Large \bf in the Field Correlator Method}

\bigskip


{Yu.A.Simonov, V.I.Shevchenko}

\bigskip

{\it  State Research
Center \\ Institute of Theoretical and Experimental Physics \\
Moscow, 117218, B.Cheremushkinskaya, 25 Russia}

\end{center}

\bigskip

\begin{abstract}

Confinement in QCD results from special properties of vacuum fluctuations of gluon fields.
There are two numerically different scales, characterizing nonperturbative
QCD vacuum dynamics: "small" one, corresponding to gluon condensate, critical temperature etc, which is about 0.1-0.3 GeV, and a "large" one, given by inverse confining string width, glueball and gluelump masses etc, which is about 1.5-2.5 GeV. We discuss this mismatch in a picture where confinement
is ensured by quadratic colorelectric field
correlators of the  special type, denoted by $D^E(x)$. The latter
together with four other quadratic correlators, $D^H, D^E_1, D^H_1$
and $D^{EH}$ support most of perturbative and nonperturbative
dynamics, while contribution of higher order terms is shown to be
small. These correlators, on the other hand, can be calculated via
gluelump Green's function, whose dynamics is defined by the
correlators themselves. In this way one obtains a self-consistent
scheme, where string tension can be expressed in terms of
$\Lambda_{QCD}$. Thus the dynamics of vacuum gluon field
fluctuations supports mean-field-type mechanism leading to
confinement at temperatures $T\leq T_c$.

\end{abstract}

\end{titlepage}






\section{Introduction}

Confinement of color is the most important property of quantum chromodynamics (QCD), ensuring
stability of  matter in the Universe. Attempts to understand physical
mechanism of confinement are incessant  since advent of constituent
quark model and QCD, for reviews see e.g. \cite{1f,2f,3f,4f}. Among
many suggestions one can distinguish three major approaches:
\begin{itemize}
\item confinement is due to classical field lumps like instantons or dyons
\item confinement can be understood
as a kind of Abelian-like phenomenon according to seminal suggestion
by 't Hooft and Mandelstam \cite{6f} of dual Meissner scenario. This
approach has gained much popularity in the lattice community and
studies of various projected objects such as Abelian monopoles and
center vortices are going on (see, e.g. \cite{5f}).
\item confinement results from properties of quantum stochastic ensemble of nonperturbative ({\it np}) fields
filling QCD vacuum. The development of this approach in a systematic
way started in 1987 \cite{7f} (see, e.g. \cite{early} as example of
earlier investigations concerning stochasticity of QCD vacuum). The
nontrivial structure of {\it np} vacuum can be described by a
set of nonlocal gauge-invariant field strength correlators (FC). We discuss this scenario in the present paper.
\end{itemize}

QCD sum rules \cite{1,2} were suggested as an independent approach to {\it np}QCD
dynamics, not addressing directly confinement mechanism. The key role is played by gluon condensate $G_2$, which is defined as {\it np} average of the following type:
\be G_2 =
\frac{\alpha_s}{\pi} \lan F_{\mu\nu}^a (0) F_{\mu\nu}^a (0) \ran \label{g2} \ee
In the sum rule framework it is a universal quantity characterizing QCD vacuum as it is, while it enters power expansion of current-current correlators with channel-dependent coefficients.
This is the cornerstone of QCD sum rules ideology.

It is worth stressing from the very beginning that $G_2$ and other condensates are usually considered as finite physical quantities. Naively in perturbation theory one would get $G_2 \sim a^{-4}$, where $a$ is a space-time ultraviolet cutoff (e.g. lattice spacing). It is always assumed that this "hard" contribution is somehow subtracted from (\ref{g2}) and the remaining finite quantity results from "soft" {\it np} fields, in some analogy with Casimir effect where modification of the vacuum by boundaries of typical size $L$ yields nonzero shift of energy-momentum tensor by the amount of order $L^{-4}$. In QCD the role of $L$ is played by dynamical scale $ \Lambda_{QCD}^{-1}$.

The idea of {\it np} gluon condensate has proved to be very fruitful.
However the relation of $G_2$ to confinement is rather tricky. Let us stress that since there can be no local gauge-invariant order parameter for confinement-deconfinement transition, $G_2$ is not an order parameter and, in particular, neither perturbative nor {\it np} contributions to this quantity vanish in either phase.\footnote{Let us repeat that following \cite{1} we define $G_2$ in (\ref{g2}) as purely {\it np} object.}
On the other hand one can show (see details in \cite{3f}) that the scale of deconfinement temperature is set
by the condensate, or, more precisely, by its electric part: $ T_c \sim G_2^{1/4}$. Let us remind
the original estimate for the condensate \cite{1}
given by $ G_2= 0.012 \>{\mbox{GeV}}^4$, with large uncertainties. One clearly sees some tension between this "small" scale and a "large" mass scale in QCD given by, e.g. the mass of the lightest $0^{++}$ glueball, which is about 1.5 GeV.

The question about the origin of this mismatch turned out to be rather deep. It was suggested in \cite{7f},
see also \cite{9f,10f} that there is another important dimensionfull parameter characterizing
{\it np} dynamics of vacuum fields: correlation length $\lambda$ (also denoted as $T_g$ in some papers).
The crucial feature of stochastic picture found in \cite{8f} is that
the lowest, quadratic, nonlocal FC, $\lan F_{\mu\nu}(x)
F_{\lambda\sigma} (y)\ran$ describes all {\it np} dynamics
with very good accuracy. It was also shown  that a simple exponential form of quadratic correlators
found on the  lattice \cite{11f} allows to  calculate all properties
of lowest mesons, glueballs, hybrids and baryons, including Regge
trajectories, lepton widths etc, see reviews \cite{2f,10f} and
references therein. Since the correlation length $\lambda$ entering these
exponents, is small, potential relativistic quantum-mechanical picture is
applicable and all QCD spectrum is defined mostly by string tension
$\sigma$ (which is an integral characteristic of the nonlocal correlator, see below) and not by
its exact profile. This is discussed in details in the next section.

However to establish the confinement mechanism unambiguously, one
should be able to calculate vacuum field distributions, i.e. field
correlators, self-consistently.  In the long run it means that
one is to demonstrate that it is essential property of QCD vacuum
fields ensemble to be characterized by correlators, which support
confinement for temperatures $T$ below some critical value $T_c$,
and deconfinement at $T>T_c$.

Attempts to achieve this goal have been undertaken in \cite{ui},
however the resulting chain of equations is too complicated to use
in practice.

 Another step in this direction is done in
\cite{12f}, where FC are calculated via gluelump  Green's functions
and self-consistency of this procedure was demonstrated for the
first time. These results were further studied and confirmed in
\cite{**}.

The main aim of this paper is to present this set of equations as a
self-consistent mechanism of confinement and to clarify qualitative
details of the FC - gluelump connection. In particular we
demonstrate how the equivalence of  colormagnetic and colorelectric
FC for $T=0$ ensures the Gromes relation \cite{13f}. We also show,
that self-consistency condition for FC as gluelump Green's function
allows to connect the mass scale to $\alpha_s$, and in this way to
express $\Lambda_{QCD}$ via string tension $\sigma$.

Lattice computations play important role as a source of independent
knowledge about vacuum field distributions. Recently a consistency
check of this picture was done on the lattice \cite{***} by
measurement of spin-dependent potentials. The resulting FC are
calculated and compared with gluelump predictions in \cite{****}
demonstrating good agreement with analytic results, in particular
small vacuum correlation length $\lambda \sim 0.1$  Fm is shown to
correspond to large gluelump mass $M_0 \approx 2$ GeV.


The paper is organized as follows. In the next section  we remind
the basic expressions for the Green's functions of $q\bar q$ and
$gg$ systems in terms of Wegner-Wilson loops and qualitative picture
of FC dynamics is discussed. Section 3 is devoted to the expressions
of spin-dependent potentials in terms of FC. We also shortly discuss
the lattice computations of FC. In section 4 FC are  expressed in
terms of gluelump Green's functions, while in section 5 the
self-consistency of resulting relations is studied. Our conclusions are presented in
section 6.

\section{Wegner-Wilson loop, Field Correlators and Green's Functions}

Since our aim  is  to study confinement for quarks (both light and
heavy) and also for massless gluons, we start with the most general
Green's functions for these objects in a proper physical background
using Fock-Feynman-Schwinger representation. The formalism is
built in such a way, that gauge invariance is manifest at all steps.
A reader familiar with this technique can skip this section and go
directly to the section 3.

First, we consider the case of mesons made of quarks while gluon
bound states are discussed below. For quarks, one forms initial and
final state operators \be \Psi_{in, out} (x,y) =\psi^\dagger (x)
\Gamma_{in,out} \Phi (x,y) \psi (y)\label{1f}\ee where $
\psi^\dagger , \psi$ are quark operators, $\Gamma$ is a product of
Dirac matrices, i.e. 1, $\gamma_\mu,\gamma_5,
(\gamma_\mu\gamma_5),...$, and $\Phi(x,y) ={ P}\exp (ig
\int^x_yA_\mu dz_\mu)$ is parallel transporter (also known as
Schwinger line or phase factor). The meson Green's function can be
written (in quenched case) as
$$G_{q\bar q} (x,y| x',y')=\lan \Psi^+_{out} (x', y') \Psi_{in}
(x,y) \ran_{A,q}=$$\be=\lan \Gamma^\dagger \Phi (x',y')
S_q(y',y)\Gamma \Phi(y,x) S_q(x,x')\ran_{A}\label{2f}\ee and the
quark Green's function $S_q$ is given in Euclidean space-time (see
\cite{15f} and references therein):
$$ S_q(x,y) = (m+\hat D)^{-1} = $$
\be = (m-\hat D) \int^\infty_0 ds (Dz)_{xy} \exp(-K)P_A\exp (ig
\int^x_y A_\mu dz_\mu) P_\sigma (x, y,s), \label{1ff} \ee where $K$
is kinetic energy term, \be K= m^2s+ \frac14 \int^s_0 d\tau
\left(\frac{dz_\mu(\tau)}{d\tau} \right)^2,\label{2ff}\ee  $m$ is
the pole mass of quark and quark trajectories $z_\mu(\tau)$ with the
end points $x$ and $y$ are being integrated over in $(D z)_{xy}$.

The factor $P_\sigma (x,y,s)$ in (\ref{1ff}) is generated by the
quark spin (color-magnetic moment) and is equal to \be
P_\sigma(x,y,s)= P_F\exp \left[g\int^s_0 \sigma_{\mu\nu} F_{\mu\nu}
(z(\tau))d\tau\right],\label{3ff}\ee where $\sigma_{\mu\nu}
=\frac{1}{4i} (\gamma_\mu\gamma_\nu-\gamma_\nu\gamma_\mu),$ and
$P_F(P_A)$ in (\ref{3ff}), (\ref{1ff}) are, respectively,  ordering
operators of matrices $F_{\mu\nu} (A_\mu)$ along the path
$z_\mu(\tau)$.

With (\ref{2f}), (\ref{1ff}) one easily gets:
$$ G_{q\bar q}(x,y;A) =$$
 \be =
\int^\infty_0 ds \int^\infty_0 ds'(Dz)_{xy} (Dz')_{xy} e^{-K-K'}
{\rm Tr }(\Gamma(m-\hat D) W_{\sigma} (x,y) \Gamma^\dagger (m'-\hat
D')),\label{5ff}\ee where Tr means trace operation both in Dirac and
color indices, while \be W_{\sigma} (x,y) = P_A \exp \left (ig
\int_{C(x,y)} A_\mu d z_\mu\right) P_\sigma (x,y,s) P'_\sigma (x,y,
s').\label{6ff}\ee In (\ref{6ff}) the closed contour $C(x,y)$ is
formed by the trajectories of quark $z_\mu(\tau)$ and antiquark
$z'_\nu(\tau')$, and the ordering $P_A$ and $P_F$ in
$P_\sigma,P'_\sigma$ is universal, i.e., $W_{\sigma}  (x,y)$  is the
Wegner--Wilson loop (W-loop) for spinor particle.

 The
factors $(m-\hat D)$ and $(m'-\hat D')$ in (\ref{5ff}) need  a
special treatment when being averaged over fields. As shown in
Appendix 1 of \cite{16*}, one can use simple replacement, \be
 m-\hat D\to  m -i \hat p, ~~ p_\mu =\frac{1}{2}
\left ( \frac{dz_\mu}{d\tau}\right)_{\tau=s}.\label{10ff} \ee

The representation Eq. (\ref{5ff}) is exact  in the limit $N_c\to \infty$,
when internal quark loops can be neglected and is a functional of
gluonic fields $A_\mu, F_{\mu\nu}$, which contains both perturbative
and {\it np} contributions, not specified at this level.

The next step is averaging over gluon fields, which yields the
physical Green's function $G_{q\bar q}$: \be G_{q\bar q} (x,y) =\lan
G_{q\bar q} (x,y;A)\ran_A.\label{7ff}\ee The averaging is done with
the usual Euclidean weight $\exp(-\mbox{Action} / \hbar)$,
containing all necessary gauge-fixing and ghost terms.

To proceed it is convenient to use nonabelian Stokes theorem (see
\cite{2f} for references and discussion) for the first factor on the
r.h.s. in (\ref{6ff}) (corresponding to the W-loop for scalar
particle) and to rewrite it as an integral over the surface $S$
spanned by the contour $C=C(x,y)$:
$$
\lan W(C)\ran = \left\lan {\rm Tr} \>P\exp\left(ig\int_C A_{\mu}(z)
dz_{\mu}\right)\;\right\ran =
 \left\lan {\rm Tr} \>{\cal P}\exp\left(ig\int_S
F(u) ds(u) \right)\;\right\ran =
$$
\be = {\rm Tr } {\cal P}_x\>\exp \>\sum\limits_{n=2}^{\infty} (ig)^n
\int_S \lan\lan F(u_{1}) .. F(u_{n}) \ran\ran d s_{1}..d s_{n} =
\exp \>\sum\limits_{n=2}^{\infty} \>{\Delta}^{(n)}[S]
\label{eq2} \ee Here $F(u_i)d s_i =
\Phi(x_0,u_i)F_{\mu\nu}^a(u_i)t^a\Phi(u_i,x_0) d s_{\mu\nu}(u_i)$
and $u_i$,  $x_0$ are the points on the surface $S$ bound by the
contour $C=\partial S$. The double brackets $\lan\lan ... \ran\ran$
stay for irreducible correlators proportional to the unit matrix in
the colour space (and therefore only spacial ordering ${\cal P}_x$
enters (\ref{eq2})). Since (\ref{eq2}) is gauge-invariant, one can
make use of generalized contour gauge \cite{gcg}, which is defined
by the condition $\Phi(x_0, u_i) \equiv 1$. Notice that throughout the paper
we normalize trace over color indices
as ${\rm Tr} {\bf 1} = 1$ in any given representation.

Since (\ref{eq2}) is an identity, the r.h.s. does not depend on the
choice of the surface, which is integrated over in $ds_{\mu\nu}
(u)$. On the other hand it is clear that each irreducible $n$-point
correlator  $\lan \lan F ... F \ran\ran$ integrated over $S$ (these
are the functions $\Delta^{(n)}[S]$ in (\ref{eq2})) depends, in
general case, on the choice one has made for $S$. Therefore it is
natural to ask the following question: is there any hierarchy of
$\Delta^{(n)}[S]$ as functions of $n$ for a given surface $S$?

To get the physical idea behind this question, let us take the limit
of small contour $C$. In this case one has for the {\it np} part of the W-loop in fundamental representation: \be - \log \lan W (C) \ran_{np} \sim  G_2 \cdot
S^2 \label{26}\ee where $S$ is the minimal area inside
the loop $C$. The {\it np} short-distance dynamics is
governed by the vacuum gluon condensate $G_2$, see (\ref{g2}).  Higher condensates and other
possible vacuum averages of local operators, in line with the Wilson
expansion, have been introduced and phenomenologically estimated in
\cite{2}, for a review see \cite{3}.

Suppose now that the size of W-loop is not small, i.e. it is larger
that some typical dynamical scale $\lambda = {\cal O}({\mbox{1 GeV}})$ to be
specified below. Let us now ask the following question: if one still
wants to expand the loop formally in terms of local condensates,
what would the structure of such expansion be? It is easy to see
from (\ref{eq2}) that there are two subseries in this expansion. The
first one is just an expansion in $n$, i.e. in the number of field
strength operators. It is just \be \lan W(C)\ran = 1+\Delta^{(2)}[S]
+ \Delta^{(3)}[S] + \frac12 (\Delta^{(2)}[S])^2 + ... \label{eq25}
\ee The second series is an expansion of each $\Delta^{(n)}[S]$ in
terms of local condensates, i.e. expansion in powers of derivatives.
Taking for simplicity the lowest $n=2$ term, this expansion reads
$$
\Delta^{(2)}[S] = \int_S d s_x \int_S ds_y \lan\lan F(x) F(y)
\ran\ran  =
$$
$$
= \int_S d s_x \int_S d s_y [ \lan\lan F(x) F(x) \ran\ran + (x -
y)_\mu \lan\lan F(x) D_\mu F(x) \ran\ran +
$$
\be
 +  (x - y)_\mu (x - y)_\nu \lan\lan
F(x) D_\nu D_\mu F(x) \ran\ran / 2 + ... ] \label{woli} \ee

It is worth mentioning (but usually skipped in QCD sum rules
analysis) that the latter series contain ambiguity related to the
choice of contours in parallel transporters. The expression
(\ref{woli}) is written for the simplest choice of straight contour
connecting the points $x$ and $y$ (and {\it not} $x$ and $x_0$ or
$y$ and $x_0$). This is the choice we adopt in what follows. In
general case each condensate in (\ref{woli}) is multiplied by some
contour-dependent function of $x$, $y$ and $x_0$. It is legitimate
to ask whether this approximation, i.e. replacement of
$$
\lan {\rm Tr}
\Phi(x_0,x)F_{\mu\nu}(x)\Phi(x,x_0)\Phi(x_0,y)F_{\rho\sigma}(y)\Phi(y,x_0)
\ran
$$
by
$$
\lan {\rm Tr} F_{\mu\nu}(x)\Phi(x,y)F_{\rho\sigma}(y)\Phi(y,x) \ran
$$
affects our final results. Since the only difference between the
above two expressions is profiles of the transporters, this is the
question about contour dependence of FCs. We will not
discuss this question in details in the present paper and refer an
interested reader to \cite{7f}. Here we only mention that the result
(weak dependence on transporters' choice) is very natural in Abelian
dominance picture since for Abelian fields the transporters exactly
cancel.

Let us stress that both expansions (\ref{eq25}) and (\ref{woli}) are
formal since the loop is assumed to be large. Moreover, from
dimensional point of view, the terms of these two expansions mix
with each other. For example, one has nonzero v.e.v. of two
operators of dimension eight: \be \Delta^{(2)}[S] \to \lan
F_{\mu\nu} D^4 F_{\mu\nu} \ran \;\; ; \;\; \Delta^{(3)}[S] \to \lan
F_{\mu\sigma} F_{\nu\sigma} D^2 F_{\mu\nu} \ran \ee with no {\it a
priori} reason\footnote{One could have such reasons if there is some
special kinematics in the problem, but this is not the case.} to
drop any of them. Moreover, even if one assumes that all these
condensates of high orders can be self-consistently defined
analogously to $G_2$, it remains to be seen whether the whole series
converges or not.

Here the physical picture behind FC method comes into play. It is assumed
(and confirmed {\it a posteriori} in several independent ways) that
if the surface $S$ is chosen as the minimal one, the dominant
contribution to (\ref{eq2}) results from v.e.v.s of the operators $F
D^k F$ and this subseries can be summed up as (\ref{woli}). In this
language nonlocal two-point FC
 can be understood as a representation for the sum of infinite sequence
 of local terms (\ref{woli}).

Physically the dominance of $\lan F D^n F \ran $ terms (and hence of
two-point FC) corresponds to the fact that the correlation
length is small yielding a small expansion parameter $\xi =
{\bar{F}}\lambda^2 \ll 1$ (${\bar F}$ is average modulus of
{\it np} vacuum fields) or, in other words, typical inverse
correlation length $\lambda^{-1}$ characterizing ensemble of QCD
vacuum fields is parametrically larger than fields themselves. The
dimensionless parameter $\xi$ in terms of condensate is given by
$({G_2}\lambda^4)^{1/2}$, and it is indeed small: of the order of
0.05 according to lattice estimates. One can say that often repeated
statement "QCD vacuum is filled by strong and strongly fluctuating
chromoelectric and chromomagnetic fields" is only partly correct -
namely, in reality the fields are not {\it so} strong as compared to
$\lambda^{-1}$. Technically one can say that we sum up the leading
subseries in (\ref{eq2}) in the same spirit as one does in covariant
perturbation theory \cite{barv} for weak but strongly varying
potentials.

Thus, for understanding of confining properties of QCD vacuum dynamics not only
scale of {\it np}  fields given by $G_2$ is important but also another
quantity - the vacuum correlation length $\lambda$, which defines
the nonlocality  of gluonic excitations. It is worth repeating here that physically
$\lambda$ encodes information about properties of the series (\ref{woli}) and has the same
theoretical status
as the corresponding {\it np} condensates in the following sense: knowledge of all terms
in (\ref{woli}) would allow to reconstruct $\lambda$. Of course, it is impossible in practice and
one has to use other methods to study large distance asymptotics of FC.
On more phenomenologic
level it was discussed  in \cite{4}, while rigorous definition was
given in the framework of the FC method\footnote{Often
referred to as Stochastic Vacuum Model} (see \cite{7f}, \cite{20*} and \cite{9f} for review).


It is important to find also the world-line representation for gluon Green's function,
which is done in the framework of background perturbation theory
(see \cite{17*} for details). In this way one finds (where $a,b$ are
adjoint color indices and dash sign denotes adjoint representation)
\be G^{ab}_{\mu\nu} (x,y) =\left\{ \int^\infty_0 ds (Dz)_{xy} e^{-K}
P_a \exp (ig \int^x_y \hat A_\mu dz_\mu) P_\Sigma
(x,y,s)\right\}^{ab}_{\mu\nu},\label{12*}\ee where
$$ P_{\Sigma} (x,y,s) = P_F \exp (2ig\int^s_0 \hat
F_{\lambda\sigma} (z(\tau)) d\tau).$$
All the above reasoning about asymptotic expansions of the corresponding W-loops is valid here as well.

Thus
the central role in the discussed method is played by quadratic
(Gaussian) FC of the form \be D_{\mu\nu,
\lambda\sigma} \equiv  g^2 \lan {\rm Tr}  F_{\mu\nu} (x) \Phi(x,y)
F_{\lambda\sigma} (y) \Phi(y, x)\ran,\label{1}\ee where $F_{\mu\nu}$
is the field strength and $\Phi(x,y)$ is  the parallel transporter.
Correlation lengths $\lambda_i$ for different channels are defined in
terms of asymptotics of (\ref{1})  at large distances:
$\exp\left(-{|x-y|}/{\lambda_i}\right)$. The physical role of
$\lambda_i$ is very  important since it distinguishes two regimes:
one expects validity of potential-type approach describing the
structure of hadrons of spatial size $R$ and at temporal scale $T_q$
for $\lambda_i\ll R, T_q$, while in the opposite case, when
$\lambda_i\gg R, T_q$ the description in terms of spatially
homogeneous condensates can be applied.

 At zero temperature the $O(4)$
invariance of Euclidean space-time holds\footnote{All treatment in
this paper as well as averaging over vacuum fields is done in the
Euclidean space. Notice that only after all Green's functions are
computed, analytic continuation to Minkowskii space-time can be
accomplished.} and FC (\ref{1}) is represented  through
two scalar functions $D(z),$ $ D_1(z)$ (where $z\equiv x-y$)  as follows
$$ D_{\mu\nu,\lambda\sigma} (z) =  (\delta_{\mu\lambda
} \delta_{\nu\sigma} -\delta_{\mu\sigma} \delta_{\nu\lambda} )D(z)
+\frac12 \left[ \frac{\partial}{\partial z_\mu} ( z_\lambda
\delta_{\nu\sigma} -z_\sigma \delta_{\nu\lambda})+\right.$$
\be\left. + \frac{\partial}{\partial z_\nu} ( z_\sigma
\delta_{\mu\lambda}
-z_\lambda\delta_{\mu\sigma})\right]D_1(z).\label{2}\ee

One has to distinguish from the very beginning perturbative and
{\it np} parts of the correlators $D(z)$, $D_1(z)$. Beginning
with the former, one easily finds at tree level \be D^{p,0}(z) = 0
\;\; ; \;\; D_1^{p,0}(z) = C_2(f)\frac{4\alpha_s}{\pi} \frac{1}{z^4}
\label{olo}\ee where $C_2(f)$ is fundamental Casimir $C_2(f) = (N_c^2 -1)/2N_c$.
At higher orders situation becomes more complicated.
Namely, one has at $n$-loop order \be D^{p,n}(z) = D_1^{p,0}(z)\cdot
G^{(n)}(z) \;\; ; \;\; D_1^{p,n}(z) = D_1^{p,0}(z)\cdot G_1^{(n)}(z)
\ee where the gauge-invariant functions $G^{(n)}(z)$, $G_1^{(n)}(z)$
has the following general structure:
$$
G^{(n)}(z), G_1^{(n)}(z) \sim \alpha_s^n \left[ c_n (\log \mu z)^n +
... \right]
$$
where $c_n$ is numerical coefficient, $\mu$ is renormalization scale
and we have omitted subleading logarithms and constant terms in the
r.h.s. Explicit expressions for the case $n=1$ can be found in
\cite{jam,ya}.

Naively one could take perturbative functions $D^{p,n}(z)$,
$D_1^{p,n}(z)$ written above and use them for computations of W-loop
or static potentials in Gaussian approximation. This however would
be incorrect. The reason is that at any given order in perturbation
expansion over $\alpha_s$ one should take into account perturbative
terms of the given order coming from all FCs, and not only
from the Gaussian one. For example, at one loop level the function
$D^{p,1}(z)$ is nonzero which naively would correspond to area law
at perturbative level. Certainly this cannot be the case, since
self-consistent renormalization program for W-loops is known not to
admit any terms of this sort \cite{23}. Technically in FC language the
correct result is restored by cancelation of contributions
proportional to $D^{p,1}(z)$ to all observables  by the terms coming
from triple FC (see details in \cite{10}).

In view of this general property of cluster expansion it is more
natural just to take relation $D^{p,n}(z)\equiv 0$ as valid at
arbitrary $n$, having in mind that proper number of terms from
(\ref{eq25}) has to be included to get this result. Thus we assume
that the following decomposition takes place: \be D(z) = D^{np}(z)
\;\; ; \;\; D_1(z) = D_1^p(z) + D_1^{np}(z) \ee and $D(z)$ has a
smooth limit when $z\to 0$. As for $D_1(z)$ its general asymptotic
at $z\to 0$ reads: \be
 D_1(z) = \frac{c}{z^4} + \frac{a_2}{z^2}
 + O(z^0) \label{12d}\ee
where $c$ and $a_2$ weakly (logarithmically) depend on $z$. Notice
that by $D_1(0)$ we always understand {\it np} "condensate" part in what follows:
\be
G_2 =  \frac{6N_c}{\pi^2} (D^{np}(0) +
D_1^{np}(0))
\label{coco}
\ee

The term $1/z^2$ deserves special consideration. One may argue, that
 at large distances the
 simple additivity of perturbative and {\it np} contributions to
 $V_{Q\bar Q}$ potential  is  violated \cite{15}.
 It is interesting whether this
 additivity holds at small distances and, in particular, mixed
 condensate $a_2$ was suggested in \cite{16} and argued to be
 phenomenologically desirable.

As we demonstrate  below our approach is self-consistent with $a_2 =
0$, i.e. when perturbative-{\it{np}} additivity holds at small
distances and there is no dimension-two condensate. However,
 at intermediate distances a complicated interrelation occurs, which does
 not exclude (\ref{12d}) being relevant in this regime.

We proceed with general analysis of (\ref{1}). As is proved in
\cite{7f},  $D(z), D_1(z)$ do not depend on relative orientation of
the plane $(\mu\nu)$, plane $(\lambda\sigma)$, and the vector
$z_\alpha$. This orientation can be of the following 3 types: a)
planes ($\mu\nu)$ and $ (\lambda\sigma)$ are perpendicular to the
vector $z_\alpha$; b) planes ($\mu\nu)$, $ (\lambda\sigma)$ are
parallel or intersecting along one direction and $z_\alpha$ lies in
one of planes; c) planes ($\mu\nu)$ and $ (\lambda\sigma)$ are
perpendicular  and $z_\alpha$ lies necessarily in one of them.

It is easy to understand that this classification is $O(4)$
invariant, and therefore one can   assign  indices $a,b,c$ to
$D_{\mu\nu, \lambda\sigma}$, resulting in  general case in three
different functions. Physically speaking the  case a) refers to e.g.
color magnetic  fields $F_{ik}, F_{lm}$ with $z_\alpha$ in the 4th
direction, and the corresponding FCs will be denoted as
$D_\bot(z)$. The case b) refers  to e.g.  the  color electric fields
$E_i(x), E_k(y)$, connected by the same temporal links along 4th
axis, and the corresponding FCs are denoted as
$D_{\parallel} (z)$. Finally, in case c) only $D_1$ part survives in
(\ref{2}) and it will be given by the  subscript EH, $D_1^{EH}$.

In case of zero temperature, when $O(4)$ invariance holds, these
three functions $D_\bot(z), D_\parallel(z), D_1^{EH}(z)$  are easily
expressed via $D(z), D_1(z)$
$$D_\bot (z)=
 D(z) +D_1(z),$$
$$ D_\parallel (z) = D(z) + D_1(z) + z^2 \frac{\partial
D_1(z)}{\partial z^2},$$
$$ D_1^{EH} (z) \equiv D_1(z),$$(note that  just $D_\bot,
D_{\parallel}$ were  measured on the lattice in \cite{11f}). For
nonzero temperature  the correlators of colorelectric and
colormagnetic fields can be different and in general there are five
FCs: $D^E(z), D_1^E(z), D^H(z), D_1^H(z)$,  and
$D_1^{EH}(z)$. As a result one obtains the full set of five
independent quadratic FCs which can be defined as follows
\be g^2 \lan {\rm Tr} H_i (x) \Phi H_j (y) \Phi^+\ran = \delta_{ij}
(D^H(z) + D_1^H (z) +\vez^2 \frac{\partial D_1^H}{\partial z^2}) -
z_iz_j\frac{\partial D_1^H}{\partial z^2},\label{3d}\ee \be g^2 \lan
{\rm Tr} E_i (x) \Phi E_j (y) \Phi^+\ran = \delta_{ij} (D^E(z) +
D_1^E (z) +z^2_4 \frac{\partial D_1^E}{\partial z^2}) +
z_iz_j\frac{\partial D_1^E}{\partial z^2},\label{4d}\ee \be g^2 \lan
{\rm Tr} H_i (x) \Phi E_j (y) \Phi^+\ran = e_{ijk} z_kz_4
\frac{\partial D_1^{EH}}{\partial z^2}.\label{5d}\ee It is
interesting that the structure (\ref{3d})-(\ref{5d}) survives
 for nonzero temperature, where $O(4)$ invariance is violated: it
 tells that  the equations (\ref{3d})-(\ref{5d}) give   the most general  form
 of the FCs. Note that at zero temperature,   $D^H= D^E,
D_1^H=D_1^E$  at coinciding point; the $O(4)$ invariance requires
that colorelectric and colormagnetic condensates coincide: $g^2\lan
{\rm Tr} H_i (x) H_i(x)\ran = g^2 \lan {\rm Tr} E_j (x) E_j(x)\ran$, hence \be
D^H(0) + D_1^H(0) = D^E(0) + D_1^E(0).\label{6}\ee Here we assume
the  {\it np} part of all FCs is finite.

At $z\neq 0$ for zero temperature FCs can be expressed
through $D_\bot (z), D_\parallel(z)$ which do not depend on whether
$E_i$ or $H_k$ enters in them, since e.g. $D_\parallel^E$ can be
transformed into $D_\parallel^H$ by  an action of $O(4)$ group
elements (the  same for $D_\bot^E, D_\bot^H$). From this one can
deduce that both $D_1$ and $D$ do not depend on subscripts $E,H$ for
zero temperature and $D_1^{EH}$ coincides with $D_1$.

Below we shall illustrate this coincidence by concrete calculations
of $D^E, D^H$ etc through the gluelump Green's functions and show
that the corresponding correlation lengths satisfy the  relations
(with obvious notations): \be \lambda^E =\lambda^H \equiv \lambda;
~~\lambda^E_1 = \lambda^H_1 = \lambda_1^{EH}\equiv
\lambda_1.\label{7}\ee As is shown below, all correlation lengths
$\lambda_j$ appear to be just inverse masses of the corresponding
gluelumps $\lambda_j = 1/M_j$.

Having analytic expressions for FCs  one might ask how to
check them versus experimental and lattice data. On experimental
side in hadron spectroscopy one measures  masses and transition
matrix elements, which are defined by
 dynamical  equation  and  the latter can be used of potential
 type due to smallness of $\lambda_j$. In static potential only
 integrals over distance enter and  the
 spin-independent static potential  can be written as   \cite{7f,9}

$$ V_{Q\bar Q} (r) = 2 \int^r_0  (r-\lambda) d\lambda
 \int^\infty_0 d\nu D^E ( \sqrt{\lambda^2 +\nu^2}) + $$
 \be +\int^r_0
 \lambda d\lambda \int^\infty_0
 D_1^E(\sqrt{\lambda^2+\nu^2}).\label{8d}\ee

 At this  point one should define how perturbative and {\it np}  contributions combine  in $D^E, D_1^E$, and this analysis was
 done in \cite{10}.
 Making use of (\ref{olo}) together with definition of the string tension  \cite{7f},
 \be \sigma^E =\frac12 \int\int d^2 z D^E(z),\label{10d}\ee
 one obtains from (\ref{8d}) the standard form of the
 static potential for  $N=3$ at distances $r\gg\lambda^E$
 \be
 V_{Q\bar Q} (r) =\sigma^E r - \frac{4\alpha_s(r)}{3 r}
 + {\cal O}(\lambda^E /r, \alpha_s^2).\label{11d}\ee
 As is argued below, $\lambda^E\approx 0.1$ Fm, so that
 the form (\ref{11d}) is applicable in the whole range of distances
 $r>0.1$ fm  provided asymptotic freedom is taken into  account in
 $\alpha_s(r)$ in (\ref{11d}).

 At the smallest values of $r$, $r\la
 \lambda^E$, one would obtain a softening of confining term,
 $\sigma r \to c r^2$ \cite{7f}, which is not seen in accurate
 lattice data at $r\ga 0.2$ fm  \cite{11,12}, imposing a stringent
 limit on the value of $\lambda^E$; $\lambda^E\la 0.1$ Fm, see
 discussion in \cite{12}.

 \section{Spin-Dependent Potentials and FC}

 We now turn to the spin-dependent interactions to demonstrate that
 FC can be extracted from them \cite{9,17,36*}.
 $V'_{1}(r),$ $ V'_{2}(r),$ $ V_{3}(r),$ $ V_{4}(r)$, plus static term   $V'_{Q\bar
 Q}(r)$ contain in integral form all five FC:
 $D^E,$ $ D_1^E,$ $ D^H,$ $ D_1^H,$ $ D_1^{HE}$ and one can extract the
 properties of the latter from the spin-dependent potentials. This procedure is used recently for comparison
 of analytic predictions \cite{****} with the lattice data \cite{***}.

 To express spin-dependent potentials in terms of FC, one can start with the W-loop (\ref{6ff}),
 entering as a kernel in the $q\bar q$
  Green's  function, where quarks $q, \bar q$ can be light or heavy. Since
   both kernels $P_\sigma, P'_\sigma$
   contain the matrix $\sigma_{\mu\nu} F_{\mu\nu}$, the  terms
   of spin-spin and spin-orbit types appear. As before, we keep Gaussian
   approximation, and in this case spin-dependent potentials
    can be computed not only for heavy, but also for light quarks. They
   have the Eichten-Feinberg form \cite{20}:
$$ V_{SD}^{(diag)}(R)=\left(\frac{\vec{\sigma}_1\vec L_1}{4\mu_1^2}-
\frac{\vec{\sigma}_2\vec
L_2}{4\mu_2^2}\right)\left(\frac{1}{R}\frac{d\varepsilon}{dR}+\frac{2dV_1(R)}{RdR}\right)+
$$
\be + \frac{\vec{\sigma}_2\vec L_1- \vec{\sigma}_1\vec
L_2}{2\mu_1\mu_2}\frac{1}{R}\frac{dV_2}{dR}+\frac{
\vec{\sigma}_1\vec {\sigma}_2V_4(R)}{12\mu_1\mu_2}+\frac{(3
\vec{\sigma}_1 \vec R\vec{\sigma}_2 \vec R-\vec{\sigma}_1
\vec{\sigma}_2 R^2)V_3} {12\mu_1\mu_2 R^2}. \label{4bb} \ee

Spin-spin interaction appear in $W_\sigma$, Eq.(\ref{6ff}) which can
be  written as
$$ \exp \left\{ -\frac{g^2}{2} \int^{s_1}_0 d\tau_1 \int^{s_2}_0
d\tau_2  \left \lan\left (
\begin{array}{ll}\vesig^{(1)} \veB& \vesig^{(1)}
\veE\\\vesig^{(1)} \veE& \vesig^{(1)} \veB\end{array}\right)_{z_1}
\left ( \begin{array}{ll} \vesig^{(2)} \veB& \vesig^{(2)}
\veE\\\vesig^{(2)} \veE& \vesig^{(2)}
\veB\end{array}\right)_{z_2}\right \ran \right\},$$ where $z_{1,2} =
z(\tau_{1,2})$ and spin-orbit interaction arises in (\ref{1}) from
the products $\lan\sigma_{\mu\nu} F_{\mu\nu} ds_{\rho\lambda}
F_{\rho\lambda}\ran$.

It is clear, that the resulting interaction will be of matrix
form $(2\times 2)\times (2\times 2)$ (not accounting for  Pauli
matrices). If  one keeps only diagonal terms in $\sigma_{\mu\nu}
F_{\mu\nu} $ (as the leading terms for large $\mu_i\approx M$) one
can write for the spin-dependent  potentials  the representation of the
Eichten-Feinberg form (\ref{4bb}).

At this point one should note that the term with
$\frac{d\varepsilon}{dR}$ in (\ref{4bb})
 was obtained from the diagonal part of the matrix  $(m-\hat D)
 \sigma_{\mu\nu} F_{\mu\nu}$, namely as product  $ i\sigma_k
 D_k\cdot \sigma_i E_i$, see \cite{9,17} for details of
 derivation, while  all other potentials $V_i, i=1,2,3,4$ are
 proportional to FCs $\lan H_i \Phi H_k \Phi\ran$. One can
 relate FCs of colorelectric and colormagnetic fields to
 $D^E, D^E_1, D^H, D_1^H$  defined by (\ref{3d})-(\ref{5d})
with the following result:
\be \frac{1}{R}\frac{dV_1}{dR}=-\int^{\infty}_{-\infty}d\nu\int^R_0
\frac{d\lambda}{R} \left (1-\frac{\lambda}{R}\right
)D^H(\lambda,\nu) \label{8} \ee \be
\frac{1}{R}\frac{dV_2}{dR}=\int^{\infty}_{-\infty}d\nu\int^R_0
\frac{\lambda d\lambda}{R^2} \left
[D^H(\lambda,\nu)+D_1^H(\lambda,\nu)+\lambda^2\frac{\partial
D_1^H}{\partial\lambda^2}\right ] \label{9} \ee \be
V_3=-\int^{\infty}_{-\infty} d\nu R^2\frac{\partial
D_1^H(R,\nu)}{\partial R^2} \label{10} \ee \be
V_4=\int^{\infty}_{-\infty}d\nu \left
(3D^H(R,\nu)+3D_1^H(R,\nu)+2R^2\frac{\partial D_1^H}{\partial
R^2}\right ) \label{11} \ee

 \be \frac{1}{R}\frac{d\varepsilon
(R)}{dR}=\int^{\infty}_{-\infty}d\nu\int^R_0 \frac{ d\lambda}{R}
\left
[D^E(\lambda,\nu)+D_1^E(\lambda,\nu)+(\lambda^2+\nu^2)\frac{\partial
D_1^E}{\partial\nu^2}\right ] \label{12} \ee

One can check, that the Gromes relation \cite{13f} acquires the form
\cite{36*}:
$$ V'_1 (R) + \varepsilon'(R) -V'_2(R)= $$ \be =
\int^\infty_{-\infty} d\nu \left [ \int^R_0 d\lambda (D^E(\lambda,
\nu)-D^H(\lambda, \nu))+\frac12 R(D_1^E(R)
-D_1^H(R))\right].\label{13}\ee

For $T=0$, when $D^E=D^H, D_1^E=D_1^H$, the Gromes relations are
satisfied identically, however for $T>0$ electric and magnetic
correlators are certainly different and Gromes relation is violated,
as one could tell beforehand, since for $T>0$ the Euclidean $O(4)$
invariance is violated.








\section{Gluelumps and FC}

In this section we establish a connection of FC $D(z)$ and $D_1(z)$
with the gluelump Green's functions.


To proceed one can use  the background field formalism
\cite{17*}\cite{22},
 where the notions of
valence gluon field $a_\mu$ and background field $B_\mu$ are
introduced, so that total gluonic field $A_\mu$
 is written as
\be A_\mu= a_\mu + B_\mu.\label{3a} \ee
 The main idea we are going to adopt here is suggested in
 the second paper of \cite{ui}. To be self-contained, let us explain it here in simple form.
 First, we single out some color index $a$ and  fix it at a given
 number. Then in color components we have, by definition,
 $$
 A_{\mu}^c = \delta^{ac} a_\mu^a + (1-\delta^{ac}) B_\mu^c
 $$
 and there is neither summation over $a$ in the first term nor over $c$ in the second.
 As a result the integration measure factorizes:
 $$
 \int \prod\limits_c {\cal D}A_\mu^c = \int {\cal D}a_\mu^a \int \prod\limits_{c\neq a}{\cal D}B_\mu^c
 $$
  so one first averages over the fields $B^c_\mu$, and after that one integrates over the fields $a^a_\mu$. The essential
 point is that the integration over
 ${\cal D}B^c_\mu$ provides colorless adjoint string attached to the gluon
 $a^a_\mu$, which keeps the color index "$a$" unchanged in the course of propagation
 of gluon $a$ in background made of $B$. This is another way of saying that gluon field of each particular color enters QCD Lagrangian quadratically. Despite such separation of the measure is a kind of trick, the formation of adjoint string is the
 basic physical mechanism  behind this background technic and it is
 related to the properties of gluon ensemble:  even for
 $N_c=3$ one has one color degree of freedom $a^a_\mu$ and 7 fields $B^c_\mu$, also
 confining string is a colorless object and therefore the singled-out adjoint index "$a$" can be preserved during interaction process  of the valence gluon $a^a_\mu$ with the background
 ($B^c_\mu)$.
These remarks to be used in what follows make explicit the notions
of the valence gluon and background field.

We assume the background Feynman gauge  $D_\mu a_\mu =0$ and assign
field transformations as follows \be a_\mu\to U^+ a_\mu U,~~
B_\mu\to U^+ (B_\mu + \frac{i}{g} \partial_\mu) U.\label{15}\ee As a
result the parallel transporter $\Phi (x,y)= P\exp ig \int^x_y B_\mu
(z) dz$ keeps its transformation property and every insertion of
$a_\mu$ between $\Phi$ transforms gauge covariantly. In what follows
we shall assume that $\Phi$ is made entirely of
the field $B_\mu$. Note that in the original cluster expansion of
the W-loop $W(B+a)$ in FC \cite{22}. \be W(B+a) =\lan {\rm Tr}
P\exp ig \int_C (B_\mu+a_\mu) dz_\mu\ran \ee
it does not matter whether $(B_\mu+a_\mu)$ or $B_\mu$ enters
$\Phi'$s since those factors cancel in the sum. Thus we assume from
the beginning that $\Phi'$s are not renormalized, since such
renormalization is unphysical anyway according to what have been
said above.


As for renormalization having physical meaning, it is known
\cite{23} to reduce to the charge renormalization and
renormalization of perimeter divergences on the contour $C$. For
static quarks the latter reduces to the mass renormalization.


In background field formalism  one has an important simplification: the combination $gB_\mu$
is renorminvariant \cite{17*,23}, so is any expression made of $B_\mu$ only.
This
point is important for comparison of FC with the lattice correlators
in \cite{***,****}. Renormalization constants for field strengths
$Z_b$ used there account for finite size of plaquettes and they are
similar to the lattice tadpole terms.

Now we turn to the analytic calculation of FC in terms of gluelump
Green's function. To this end we insert (\ref{3a}) into (\ref{1}) and
have for $F_{\mu\nu}(x)$: \be F_{\mu\nu} (x) =\partial_\mu
A_\nu-\partial_\nu A_\mu-ig [A_\mu,A_\nu] =\hat D_\mu a_\nu-\hat
D_\nu a_\mu-ig[a_\mu,a_\nu]+ F^{(B)}_{\mu\nu}.\ee Here, the term
$F_{\mu\nu}^{(B)}$ contains only the field $B_\mu^b$. It is clear
that when one averages over field $a_\mu^a $ and sums finally over
all color indices  $a$, one actually exploits all the fields with
color indices from $F^{(B)}_{\mu\nu}$, so that the term
$F_{\mu\nu}^{(B)}$ can be omitted, if summing over all  indices $a$
is presumed to be done at the end of calculation.

 As a result $D_{\mu\nu,\lambda\sigma}$ can be
written as \be D_{\mu\nu,\lambda\sigma} (x,y) =
D^{(0)}_{\mu\nu,\lambda\sigma} + D^{(1)}_{\mu\nu,\lambda\sigma}
+D^{(2)}_{\mu\nu,\lambda\sigma}\label{18}\ee where the superscript
0,1,2 refers to the power of $g$ coming from the term $ig [a_\mu,
a_\nu]$.

We can address now an important question about the relation between
FCs and gluelump Green's functions. We begin with
1-gluon gluelump, whose Green's function reads \be G^{(1g)}_{\mu\nu}
(x,y) = \lan {\rm Tr}_a a_\mu(x) \hat \Phi (x,y) a_\nu
(y)\ran.\label{20}\ee According to (\ref{15})
$G^{(1g)}_{\mu\nu}(x,y)$ is a gauge invariant function.

As shown in \cite{12f}, the first term in (\ref{18}) is connected to
the functions $D_1^E, D_1^H$ and it can be written as follows\be
D_{\mu\nu,\lambda\sigma}^{(0)} (x,y) =\frac{g^2}{2N^2_c}\left\{
\frac{\partial}{\partial x_\mu}\frac{\partial}{\partial y_\lambda}
G^{(1g)}_{\nu\sigma}(x,y) + perm\right\} +
\Delta^{(0)}_{\mu\nu,\lambda\sigma},\label{19}\ee where
$\Delta^{(0)}_{\mu\nu,\lambda\sigma}$ contains contribution of
higher FCs, which we systematically discard.

On the other hand one can find $G^{(1g)}_{\mu\nu}(x,y)$ from the
 expression \cite{15f} written as \be G^{(1gl)}_{\mu\nu} (x,y) = {\rm Tr}_a
\int^\infty_0 ds (Dz)_{xy} \exp(-K) \lan W^F_{\mu\nu}
(C_{xy})\ran,\label{21}\ee where the spin-dependent W-loop is
\be W^F_{\mu\nu} (C_{xy})= PP_F \left\{ \exp (ig \int B_\lambda
dz_\lambda) \exp F\right\}_{\mu\nu}\label{22}\ee and the  gluon spin
factor is $ \exp  F \equiv \exp (2ig \int^s_0 d\tau \hat F_B
(z(\tau)))$ with $\hat F_B$ made of the background field $B_\mu$
only.

Analogous expression can be constructed for Green's function of
2-gluon gluelump. It is given by the following expression
$$ G_{\mu\nu, \lambda\sigma}^{(2gl)}(x,y) =\lan {\rm Tr}_a (f^{abc}f^{def} a_\mu^a
(x) a^b_\nu (x) T^c \hat \Phi (x,y) T^f\times $$ \be\times
a^d_\lambda (y) a^e_\sigma(y) \ran \equiv N^2_c (N^2_c -1)
(\delta_{\mu\lambda} \delta_{\nu\sigma} - \delta_{\mu\sigma}
\delta_{\nu\lambda}) G^{(2gl)}(z).\label{32}\ee At small distances
$G^{(2gl)}(x,y)$ is dominated by perturbative expansion terms, \be
G^{(2gl)} (z) \sim \frac{1}{z^4} ,\label{33}\ee however,
as we already discussed in details, all these perturbative terms are
canceled by those from higher FCs (triple, quartic, etc...),
therefore expansion in fact starts with {\it np} terms of dimension four.

To identify the {\it np} contribution to $G^{(2gl)}(z)$ we
rewrite it as follows: \be G^{(2gl)}(z) =\int^\infty_0 ds_1
\int^\infty_0 ds_2 (Dz_1)_{0x} (Dz_2)_{0x} {\rm Tr} W_\Sigma (C_1,
C_2)\label{48*}\ee where the two-gluon gluelump W-loop
$W_\Sigma (C_1, C_2)$ is depicted in Fig. 6 and can be written as
(in the Gaussian approximation) \be {\rm Tr} W_\Sigma(C_1, C_2) =\exp
\left\{ -\frac12 \int_s\int d\pi_{\mu\nu} (u) d\pi_{\lambda\sigma}
(v) \lan \hat F_{\mu\nu} (u) \hat \Phi\hat F_{\lambda\sigma} (v)\ran
\right\}.\label{49}\ee and the total surface $S$ consists of 3
pieces, as shown in Fig.7 \be F_{\mu\nu} d\pi_{\mu\nu} (u) =
F_{\mu\nu} d s_{\mu\nu} (u) -2 ig  d\tau (\hat F(u))\label{50*}\ee
where $(\hat F)$ has Lorentz tensor and adjoint color indices
$F^{ab}_{ij}$ and lives on gluon trajectories, i.e. on the
boundaries of $S_i$. In full analogy with (\ref{19}) we have \be
D(z) = \frac{g^4 (N_c^2-1)}{2} G^{(2gl)} (z)\label{313}\ee

The crucial point is that the Green's function (\ref{19}),
(\ref{48*}) can be calculated in terms of the same FCs
$D(z)$, $D_1(z)$. Indeed one has  \be G^{(w)}(z) = \lan f |
\exp(-H_w |z|) | i\ran \label{nek} \ee where the index $w$ stays for
1-gluon or 2-gluon gluelump Hamiltonians. The latter are expressed
via the same FCs $D(z)$, $D_1(z)$ (see \cite{19} and references therein):
\be H_w = H_0[\mu] + \Delta H_L[\mu, \nu] + \Delta H_{Coul}[D_1] +
\Delta H_{string}[D,\nu]
 \ee
 where the last term $H_{string}[D, \nu]$ depends on $D(z)$ via
 adjoint string tension
 \be
 \sigma_{adj} = \frac94 \int_0^{\infty} d^2 z D(z)
 \ee
and Hamiltonian depends on einbein fields $\mu$ and $\nu$. The term
corresponding to perimeter Coulomb-like interaction $\Delta
H_{Coul}[D_1]$ depends on $D_1(z)$ (compare with (\ref{8d}),
(\ref{4bb})).

The self-consistent regimes  correspond to different asymptotics of
the solutions to these equations. In Coulomb phase of a gauge theory
both $H_{1g}$, $H_{2g}$ exhibit no mass gap, i.e. large $z$
asymptotic of (\ref{nek}) is power-like. The function $D(z)$
vanishes in this phase and W-loop obeys perimeter law.
In the confinement phase realized in Yang-Mills
theory at low temperatures a typical large-$z$ pattern is given by
$$
D(z), D_1(z) \sim \exp(-|z|/\lambda_i)
$$
i.e. there is a mass gap for both $H_{1g}$, $H_{2g}$.

Confining solutions are characterized by W-loops obeying area law.
In other words, Hamiltonians expressed in terms of interaction
kernels depending on $D(z)$, $D_1(z)$ exhibit mass gap if these
kernels are confining. On the other hand the same mass gap plays a
role of inverse correlation length of the vacuum. This should be
compared with well known mean-field technique. In our case the role
of mean-field is played by quadratic FC, which
develops non-trivial Gaussian term $D(z)$. Notice that the
exponential form of its large distance asymptotics $\exp(-|z|/\lambda)$ (and not,
let's say, $\exp(-z^2/\lambda^2)$ ) is dictated by spectral expansion
of the corresponding Green's function at large distances.

Full solution of the above equations is a formidable task not
addressed by us here. Instead, as a necessary prerequisite we check
below different asymptotic regimes and demonstrate self-consistency
of the whole picture.

We begin with small distance region. The {\it np} part of contribution to
$D_1(z)$ at small distances comes from two possible  sources: the
area law term (first exponent and the $\exp F$ term
in (\ref{22})). Both contributions are depicted in Fig.1 and Fig.2
respectively. We shall disregard the term $\exp F$ in this case,
since for the one-gluon gluelump it does not produce hyperfine
interaction and only gives rise to the {\it np} shift of the gluon mass,
which anyhow is eliminated by the renormalization (see Appendix  for
more detail). This is in contrast to the two-gluon gluelump Green's
function generating $D^E, D^H$, where the hyperfine interaction
between two gluons is dominating at small distances.

As a result Eq. (\ref{20}) without the ($\exp F$) term yields \be
D_1(z) = \frac{4C_{2}(f)\alpha_s}{\pi} \frac{1}{z^4} +\frac{g^2}{12}
G_2.\label{27}\ee  It is remarkable that the sign of the {\it np} correction is
positive.
%

 At large distances one can use the gluelump
Hamiltonian for one-gluon gluelump from \cite{24} to derive the
asymptotics \cite{12f} \be D_1(z) = \frac{2C_2(f) \alpha_s M_0^{(1)}
\sigma_{adj}}{|z|} e^{-M_0^{(1)}|z|},~~ |z| M_0^{(1)} \gg1.\label{30}\ee
 where $M_0^{(1)} = (1.2 \div 1.4)$ GeV for
$\sigma_f =0.18$ GeV$^2$ \cite{24,25}.

We now turn to the FC $D(z)$ as was studied in \cite{12f}.
  The relation (\ref{313})
 connects $D(z)$ to the two-gluon gluelump Green's function,
 studied in \cite{24} at large distances. Here we need its small -
 $z$ behaviour and we shall write it in the form
\be G^{(2gl)} (z) = G_{p}^{(2gl)} (z) + G^{(2gl)}_{np}(z)
\label{47*}\ee where $G^{(2gl)}_{p} (z)$
contains purely perturbative contributions which are subtracted  by
higher-order FCs,  while $G^{(2gl)}_{np}(z)$ contains
{\it np} and possible perturbative - {\it np}
interference terms. We are interested in the contribution of the
FC $\lan FF\ran$ to $G^{(2gl)}(z)$, when $z$ tends to zero.

One can envisage three types of contributions:

a) due to product of surface elements $ds_{\mu\nu}
ds_{\lambda\sigma}$, which gives for small surface the factor
$\exp\left(-\frac{g^2\lan FF\ran S^2_i}{24 N_c}\right)$, similar to
the situation discussed for $D_1$. This is depicted in Fig.7.

b) contribution of the type $d\tau_1 d\tau_2\lan F F\ran$, where
$d\tau$ and $d\tau_2$ belong to different gluon trajectories, 1 and
2, as depicted in Fig.8. This is the hyperfine gluon-gluon
interaction, taken into account in \cite{24} in the course of the
gluelump mass calculations (regime of large $z$), however in that
case mostly the perturbative part of $\lan FF\ran$ contributes (due
to $D_1$). Here we keep in $\lan FF \ran$ only the $NP$ part and
consider the case of small $z$.

c) Again the term $d\tau^i d\tau_j$, but now $i=j$. This is actually
a part of the gluon selfenergy correction, which should be
renormalized to zero, when all (divergent) perturbative
contributions are added. As we argued in Appendix of \cite{12f}, we
disregard this contribution here, as well as in the case of $D_1(z)$
(one-gluon gluelump case). Now we treat both contributions a) and
b).

a) In line with the treatment of $D_1$, Eq. (\ref{27}), one can
write the term, representing two-gluon gluelump as two nearby
one-gluon gluelumps
which yields for $D(z)$, \be \Delta D_a(z)
=-\frac{g^4N_cG_2}{4\pi^2}\label{62*}\ee b) in this case one should
consider the diagram given in Fig. 8, which yields the answer (for
details see Appendix).

\be G_b^{(2gl)} (z) = \frac{4N_c^2}{N_c^2-1} \int \frac{d^4 wd^4 w'
D(w-w')}{(4\pi^2)^4 (w-z)^2 w^2(w'-z)^2 w^{'2}}\label{52*} \ee which
contributes to $D(z)$ as \be \Delta D_b(z) =2N_c^2 g^4
h(z).\label{64*}\ee where at small $z$, $h(z) \approx
\frac{D(\lambda_0)}{64\pi^4} \log^2 \left(
\frac{\lambda_0\sqrt{e}}{z}\right)$, $\lambda_0 $ is of the order of
correlation length $\lambda$.

At large distances one use the two-gluon gluelump Hamiltonian as in
\cite{25} and find the corresponding spectrum and wave functions,
see \cite{25} and appendix 5 of \cite{12f} for details. As a result
one obtains in this approximation \be D(z) =\frac{g^4(N^2_c-1)}{2}
0.1 \sigma^2_f e^{-M_0^{(2)} |z|},~~ M_0^{(2)}
|z|\gg1\label{34}\ee where $M^{(2)}_0 $ is the lowest two-gluon
gluelump mass found in \cite{24} to be about $ M_0^{(2)} = (2.5 \div
2.6)$  GeV

We shall discuss the resulting properties of $D(x)$ and $D_1(x)$ in
the next section.

\section{Discussion of Consistency}

We start with the check consistency for $D(z)$. As is shown above,
$D(z)$  has the following behavior at small $z$ \be D(z) \approx
- 4 N_c
\alpha^2_s (\mu(z)) G_2 +
N_c^2 \frac{\alpha_s^2(\mu(z))}{2\pi^2} D(\lambda_0) \log^2 \left(
\frac{\lambda_0\sqrt{e}}{z}\right)
\label{36aa}
\ee
Since $\alpha_s(\mu(z))\sim 2\pi/\beta_0 \log (\Lambda z)^{-1}$
the first term is subleading at $z\to 0$ and the last term on the
r.h.s. tends to a constant:
\be
D(0) = \frac{N^2_c}{2\pi^2 }D(\lambda_0)
\left(\frac{2\pi}{\beta_0}\right)^2\label{36aa1} \ee Since from (\ref{36aa1})
 one can infer, that $D(0)\approx 0.15 D(\lambda_0)$ for $N_c=3,$ where
$\lambda_0 \ga \lambda^E$. So $D(z)$ is an increasing function of
$z$ at small $z$, $ z \la \lambda_0$,  and for $z\gg \lambda$ one
observes exponential falloff. The qualitative picture illustrating this solution for $D(z)$ is shown in Fig.5.

This pattern may solve qualitatively the contradiction
between the values of $D(0)$ estimated from the string tension
$D_\sigma(0)\simeq\frac{\sigma}{\pi\lambda^2}\approx 0.35$ GeV$^4$
and the value obtained in naive way from the gluon condensate $D_{G_2} (0)
=\frac{\pi^2}{18} G_2\approx (0.007\div 0.012)$ GeV$^4$. One can
see that $D_\sigma(0) \approx (30\div 54) D_{G_2} (0)$. This seems to be a reasonable explanation of the mismatch discussed in the introduction.

As was shown in \cite{12f}, the large distance exponential behavior
is selfconsistent, since (assuming that it persists for all $z$,
while small $z$ region contributes very little) from the equality
$\sigma = \pi\lambda^2 D_{\sigma}(0)$, comparing with (\ref{34}) one obtains  \be 0.1\cdot 8\pi^2
\alpha^2_s (N_c^2-1)
\sigma^2_f=\frac{\sigma_f}{\pi\lambda^2}\label{72}\ee
 where in $\alpha_s(\mu)$ the scale $\mu$ corresponds roughly to
 the gluelump average momentum (inverse radius) $\mu_0\approx 1$ GeV. Thus (\ref{72}) yields $\alpha_s
 (\mu_0) \approx 0.4$ which is in reasonable agreement with $\alpha_s$ from other systems \cite{bm}.

 We end this section with discussion of
three points:

 1) $D(z) $ and $D_1(z)$ have been obtained here
in the leading approximation, when gluelumps of minimal number of
gluons contribute: 2 for $D(z)$ and 1 for $D_1(z)$. In the higher
orders of $O(\alpha_s)$ one has an expansion of the type
$$ D(z) = D^{(2gl)} (z) + c_1 \alpha_s^3 D^{(1gl)} (z) + c_2
\alpha^3_s D^{(3gl)} (z) +...$$ \be D_1(z) =D^{(1gl)} (z) + c'_1
\alpha^3_s D^{(2gl)} (z) +...\label{36}\ee Hence the asymptotic
behavior for $D(z)$ will contain exponent of $M_0^{(1)} |z|$ too,
but with a small  preexponent coefficient.

2) The behavior of $D(z), D_1^{(np)}(z)$ at small $z$ is defined by
NP terms of dimension four, which are condensate $G_2$ as in
(\ref{26}), and the similar term from  the expansion of $\exp F$,
namely, \be \lan \exp F\ran =1+ 4g^2 \int^s_0 d\tau \int^{s'}_0
d\tau' \lan  F (u(\tau)) F(u(\tau')\ran +...\label{37}\ee therefore
one does not encounter mixed terms like $O(\frac{m^2}{z^2}),$
however if one assumes that expansion of $\lan F(0) F(z)\ran$ starts
with terms of this sort as it was suggested  in \cite{16}, then one
will have a selfconsistent
condition for the coefficient in front of this term.

3) To study the difference between $D^E_1, D_1^H$ at $T\leq T_c$ one
should look at Eq. (\ref{19}) and compare the situation, when
$\mu=\lambda=4, ~~ \nu=\sigma =i$ and take $z=x-y$ along the 4-th
axis (for $D_1^E$), while for $D_1^H$ one takes $\mu=\lambda=i$,
$\nu=\sigma=k$ and the same $z$. One can see that in both cases one
ends up with the one-gluon gluelump Green's function $G_{\mu\nu}$
(\ref{21}) which in the lowest (Gaussian) approximation is the same
$G_{\mu\nu} =\delta_{\mu\nu} f(z)$, and $f(z)$ is the standard
lowest mass (and lowest angular momentum) Green's function.

Hence $D_1^E=D_1^H$ in this approximation and for $T=0$ this is an
exact relation, as was discussed above. Therefore \be
\lambda_1=\lambda_1^E =\lambda_1^H=1/M_0^{(1)},~~ M_0^{(1)} \approx
1.2\div 1.4~{\rm GeV},~~ \lambda_1\cong 0.2 \div 0.15 ~{\rm
fm}.\label{38}\ee The value $M_0^{(1)}$ in (\ref{38}) is taken from
the calculations in \cite{24}. The same is true for $D^E,D^H$, as it
is seen from (\ref{32}), where $G^{(2gl)}$ corresponds to
 two-gluon subsystem angular momentum $L=0$ independently of $\mu\nu,
\lambda\sigma$.

Hence one obtains \be \lambda\equiv \lambda^E=\lambda^H
=\frac{1}{M^{(2)}_0};~~ \lambda \cong 0.08~{\rm fm},~~ M_0^{(2)}
\approx 2.5 ~{\rm GeV}.\label{39}\ee where the value $M_0^{(2)}\approx
2.5$ GeV is taken from \cite{24}.

\section{Conclusions}

We have derived, following the method of \cite{17*} the expressions for
FC $D(z)$, $D_1(z)$ in terms of gluelump Green's functions. This is
done in Gaussian approximation. The latter are calculated using
Hamiltonian where {\it np} dynamics is given by $D^{np}(z)$,
$D_1^{np}(z)$. In this way one obtains selfcoupled equations for
these functions, which allow two types of solutions: 1)
$D^{np}(z)=0$, $D_1^{np}(z)=0$, i.e. no {\it np} effects at
all; 2) $D^{np}(z)$, $D_1^{np}(z)$ are nonzero and defined by the
only scale, which should be given in QCD, e.g. string tension
$\sigma$ or $\Lambda_{QCD}$. All other quantities are defined in
terms of these basic ones. We have checked consistency of
selfcoupled equations at large and small distances and found that to
the order ${\cal O}(\alpha_s)$ no mixed perturbative-{\it{np}}
terms appear. The function $D_1(z)$ can be represented as a sum
of perturbative and {\it np} terms, while $D(z)$ contains
only {\it np} contributions.

We have found a possible way to explain the discrepancy between the
average values of field strength taken from $G_2$ and from $\sigma$
by showing that $D(z)$ has a local minimum at $z=0$ and grows at
$z\sim \lambda$. Small value of $\lambda$ and large value of the
gluelump mass $M_{gl}=1/\lambda \approx 2.5$ GeV explains the
lattice data for $\lambda \approx 0.1$ Fm. Thus the present
paper argues that relevant degrees of freedom ensuring confinement
are gluelumps, described self-consistently in the language of FCs.

\bigskip

{\bf Acknowledgments }

\medskip

The work is
supported by the grant for support of scientific schools
NS-4961.2008.2. The work of V.S. is also supported by INTAS-CERN
fellowship 06-1000014-6576.

\appendix

\section{Nonperturbative contributions to  $D_1(z)$ and $D(z)$ at small $z$ }

We start with $D_1(z) $ expressed via  gluelump Green's function
(\ref{20}), (\ref{21})  $G_{\mu\nu} (z) = \delta_{\mu\nu} f(z)$.

The leading $NP$ behavior at small $z$, proportional to $\lan F(x)
F(y)\ran$, is obtained from the amplitudes, shown in Fig.1 and
Fig.2. For the amplitude of Fig.1 one can use the
 vacuum average of the W-loop (\ref{26}) in the limit of
 small contours and neglecting the factor $\exp F$ in (\ref{22}).
 One obtains for the adjoint loop
 \be \lan W(C_{xy})\ran =\exp \left( - \frac{g^2\lan
 (F^a_{\mu\nu}(0)^2\ran}{24 N_c} \gamma S^2\right)\label{A1}\ee
 where
 $\gamma=\frac{C_2(adj)}{C_2(f)}= \frac{2N^2_c}{N^2_c-1}$
 (original derivation in \cite{12f} referred to the fundamental
 loop). Proceeding as in Appendix 2 of \cite{12f} one obtains the
 $NP$ correction  $\Delta D_1$
\be\Delta D_{1a}=C_2 (adj) \frac{g^2}{12N_c} G_2= \frac{g^2}{12}
 G_2\label{A2}\ee
 where we have introduced the standard definition \cite{1},
 $G_2\equiv \frac{\alpha_s}{\pi} \lan F^a_{\mu\nu} (0)
 F^a_{\mu\nu} (0)\ran$. One can check, that $\Delta D_1 ={\cal O}
 (N^0_c)$.

 We turn now to the amplitude of the Fig.2. The corresponding
 $f(x)$ for $x\to 0$ can be written as follows \be \Delta
 f_{1b}(x) =\int d^4 u d^4v G(0,u) G(u,v) G(v, x) 4\pi^2 N^2_c
 G_2\label{A3}\ee where the gluon Green's function $G(x,y)$ for
 small $|x-y|$ can be replaced by the perturbative part: $G(x,y)
 \to G_0 (x,y) =\frac{1}{4\pi^2 (x-y)^2}$.

 Taking into account that $\Delta D_1(x) =- \frac{2g^2}{N^2_c}
 \frac{d}{dx^2} \Delta f(x)$, one obtains for $x\to 0$,
 \be
\Delta D_{1b} (x) = 2 g^2 G_2 I(x)\label{A4}\ee where we have
defined \be I(x) =-\frac{d}{dx^2} \int \frac{d^4u d^4v}{(4 \pi^2)^2
u^2 (u-v)^2 (v-x)^2}.\label{A5}\ee

The integral $I(0)$ diverges at small $v$  and large $u,v$. The
latter divergence is removed since at large arguments $G(x,y)$ is
damped by confining force, keeping the propagating gluon nearby the
4-th axis, $x_4=y_4=0$ where the static gluonic source resides. One
can estimate $ I(x\to 0)\sim {\cal O}((ln \frac{\lambda^2}{x^2})^\alpha))$.
Since $G_2$ is connected to $D, D_1$ \be G_2 = \frac{3N_c}{\pi^2}
(D^E(0) + D^E_1(0) + D^H(0)+D^H_1(0))\label{A6}\ee one can see in
(\ref{A4}), that coefficients of $D^E_1$ on both sides of (\ref{A4})
have the same order of magnitude and the same sign, suggesting a
selfconsistency on this preliminary level. However, as we argued
before and in \cite{12f} this contribution is actually gluon mass
renormalization, which is zero. This is especially clear when $\lan
F F\ran$ in Fig.2 is replaced by its perturbative part.

We turn now to the most important case of the confining FC
$D^E(x)$. The corresponding two-gluon gluelump amplitude is depicted
in Fig.3, and the $np$ contributions to $D^E(x)$ are given in Fig.4
and Fig.5.

We start with the amplitude of Fig.4, which can be represented as a
doubled diagram of Fig.1 and therefore the contribution to the
two-gluon gluelump Green's function $G^{(2gl)} (x)$ will be \be
G^{(2gl)} (x) =\frac{1}{(4\pi^2 x^2)^2}-\frac{\gamma G_2}{24
N_c (4\pi^2)}.\label{A7}\ee Now using relation (\ref{26}) one
obtains for the {\it np} contribution of Fig.4 \be \Delta D_a (x)
=\frac{ g^4 (N^2_c-1)}{2} \Delta G^{(2gl)} (x) =-
\frac{g^4N_c}{4\pi^2} G_2.\label{A8}\ee Note, that this contribution
is finite at $x\to 0$, however with the negative sign. We now turn
to the {\it np} contribution of Fig.5, which can be obtained from
(\ref{32}) inserting there the product of operators $\hat F_{fg} (w)
d^4 w \hat F_{f'g'} (w') d^4 w'$, where $fg (f'g')$ are adjoint
color indices, $\hat F_{fg} = F^a T^a_{fg} =F^a(-i) f_{a fg}$. Using
the formulas \be f_{abc} f_{a'bc} = N_c \delta_{aa'};~~ f_{abc}
f_{ade} f_{bdf} f_{ceg} = \frac{N^2_c}{2} \delta_{fg},\label{A9}\ee
one arrives at the expression \be \Delta D_{b} (x) = 2 N^2_c g^4
h(x),\label{A10}\ee $$ h(x) =\int {d^4wd^4w'} D(w-w') G(w) G(w')
G(w-x) G(w'-x)$$ where  $G(y)$ is the  gluon Green's function in the
two-gluon gluelump; the gluon is connected at large distances by two
strings to another gluon and to the static gluon source, see Fig.5.
At small $x$ (we take it along axis 4 for convenience) the integral
in $h(x)$ grows when $G(y)$ is close to the 4th axis and becomes the
free gluon, $G(y) \to G_0(y) =\frac{1}{4\pi^2 y^2}$ . As a result
one obtains \be h(x) \to h_0(x) =\int \frac{d^4 w d^4
w'}{(4\pi^2)^4} \frac{D(w-w')}{w^2w^{'2}(w-x)^2
(w'-x)^2}\label{A11}\ee and one should have in mind that the
integral for $x\to 0$ diverges both at small and at large $w, w'$.
The small $w, w'$ region will give the terms $O(ln
\frac{\lambda^2_0}{x^2})$, and at large $w,w'$ the integral is
protected by the fall-off  of $G(y)$ at large $y$ due to
confinement, therefore we must imply in  the integrals in
(\ref{A11}) the upper limits for $w,w'$ at some $\lambda_0 \ga
\frac{1}{\sqrt{\sigma_{adj}}}$. Then introducing for $w(\vew, w_4)$
polar coordinates $|\vew|=\rho \sin \theta, w_4 =\rho\cos \theta$
(and the same for $w'$), one arrives at the integral \be \int^\pi_0
\frac{\sin^2 \theta d\theta}{\rho^2 - 2x \rho \cos \theta
+x^2}=\frac{\pi}{2\rho^2} (\rho\geq x) \label{A12}\ee or
$\frac{\pi}{2x^2}(x\geq \rho)$, and finally one has the estimate of
contribution to (\ref{A10}) from the region of small $w, w' (w|w|,
|w'|\la \lambda_0$) \be h(x) \approx
\frac{D(\lambda_0)}{4(4\pi^2)^2} {\log}^2
\left(\frac{\lambda_0\sqrt{e}}{x}\right)\label{A13}\ee and finally
for $D(x)$ one obtains (omitting the perturbative term in
(\ref{A7}), which cancels, \be D(x) =-
\frac{g^4 N_c}{4\pi^2} G_2 +\frac{N^2_c}{2\pi^2} \alpha^2_s
D(\lambda_0)
{\log}^2\left(\frac{\lambda_0\sqrt{e}}{x}\right).\label{A14}\ee

Note, that the last term on the r.h.s. dominates at small $x$ and
compensates the decrease of the $\alpha^2_s$ term, since \be
\alpha^2_s (x) ln^2
\left(\frac{\lambda_0\sqrt{e}}{x}\right)\cong\left(\frac{4\pi ln
\frac{\lambda_0\sqrt{e}}{x}}{\beta_0 ln \left(\frac{1}{x^2
\Lambda^2}\right)}\right) \to \left(\frac{4\pi}{\beta_0}\right)^2
(x\to 0)\label{A15}\ee Hence \be D(x<\lambda_0) \approx
\frac{N^2_c}{2\pi^2}\left(\frac{4\pi}{\beta_0}\right)^2D(\lambda_0)
< D(\lambda_0)\label{A16}\ee


\newpage

\begin{figure}
\begin{center}
 \includegraphics[width=45mm,keepaspectratio=true]{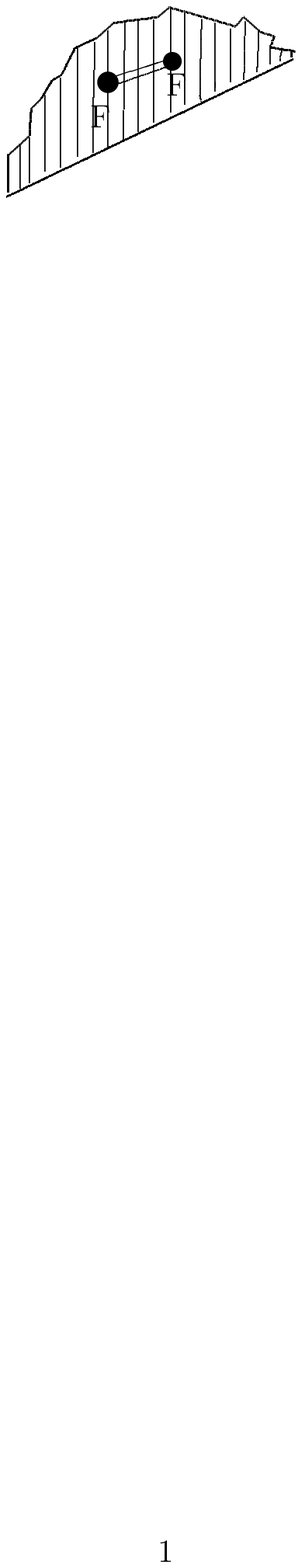}
\end{center}
\centerline{Fig. 1}
\end{figure}
\begin{figure}
\begin{center}
 \includegraphics[width=45mm,keepaspectratio=true]{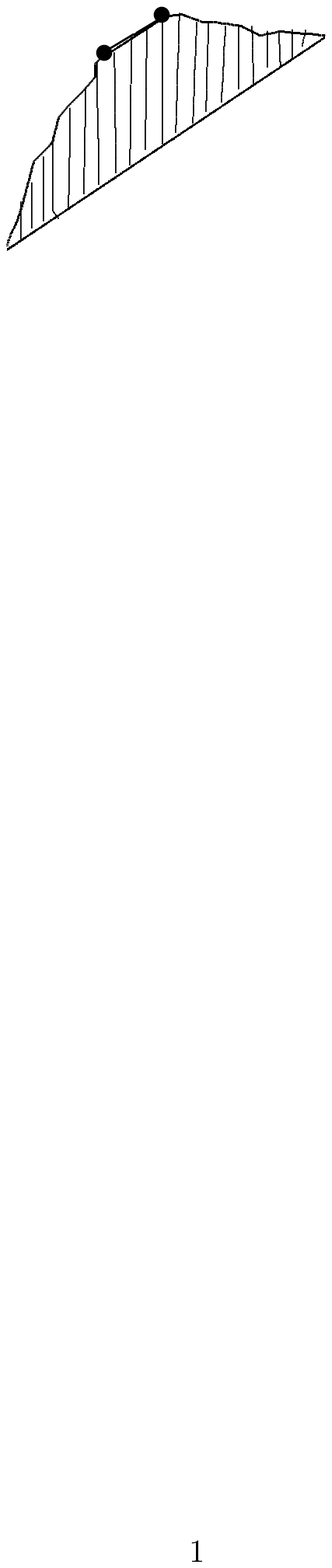}
\end{center}
\centerline{Fig. 2}
\end{figure}
\begin{figure}
\begin{center}
 \includegraphics[width=45mm,keepaspectratio=true]{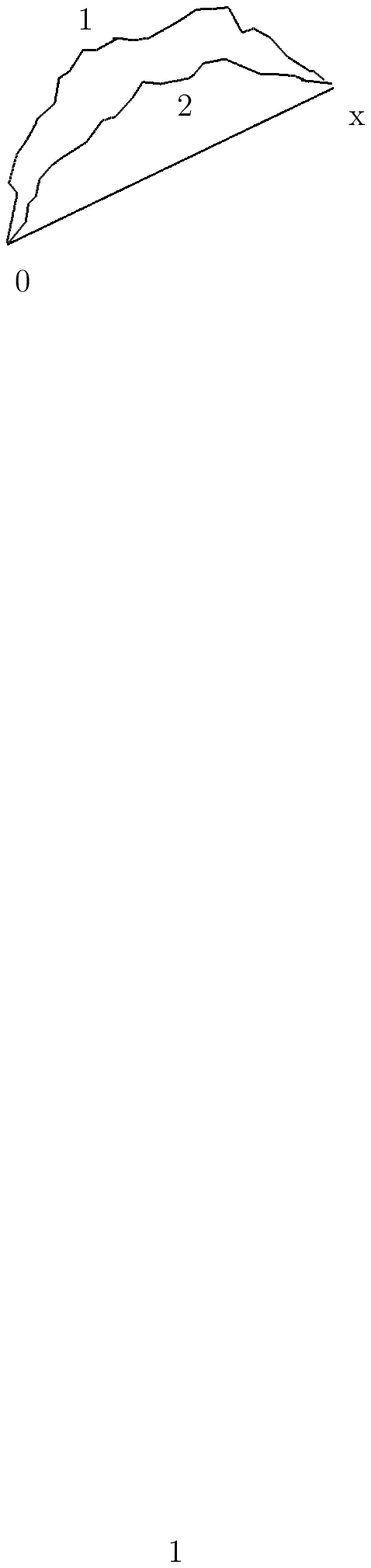}
\end{center}
\centerline{Fig. 3}
\end{figure}
\begin{figure}
\begin{center}
\includegraphics[width=45mm,keepaspectratio=true]{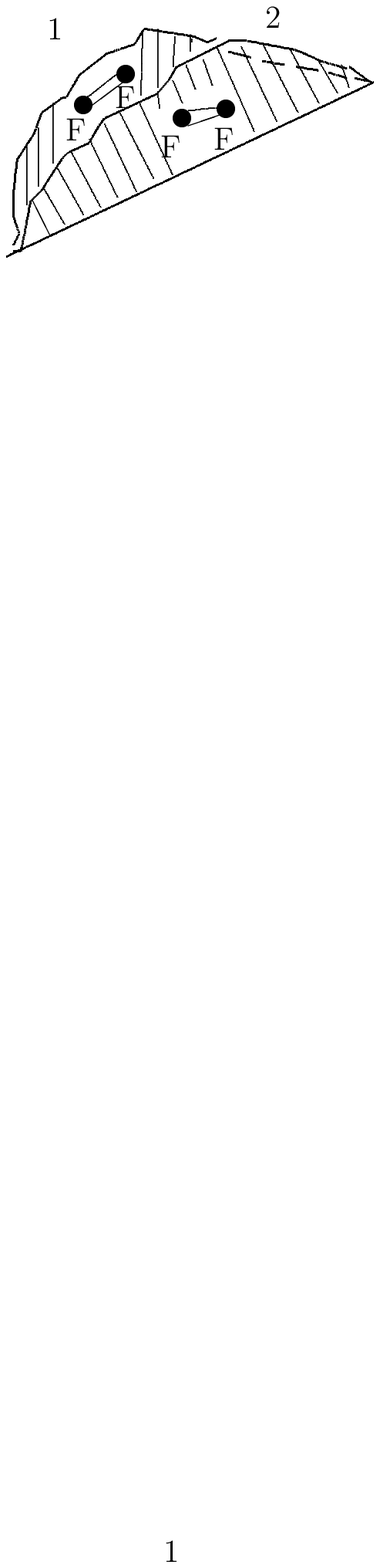}
\end{center}
\centerline{Fig. 4}
\end{figure}
\begin{figure}
\begin{center}
 \includegraphics[width=45mm,keepaspectratio=true]{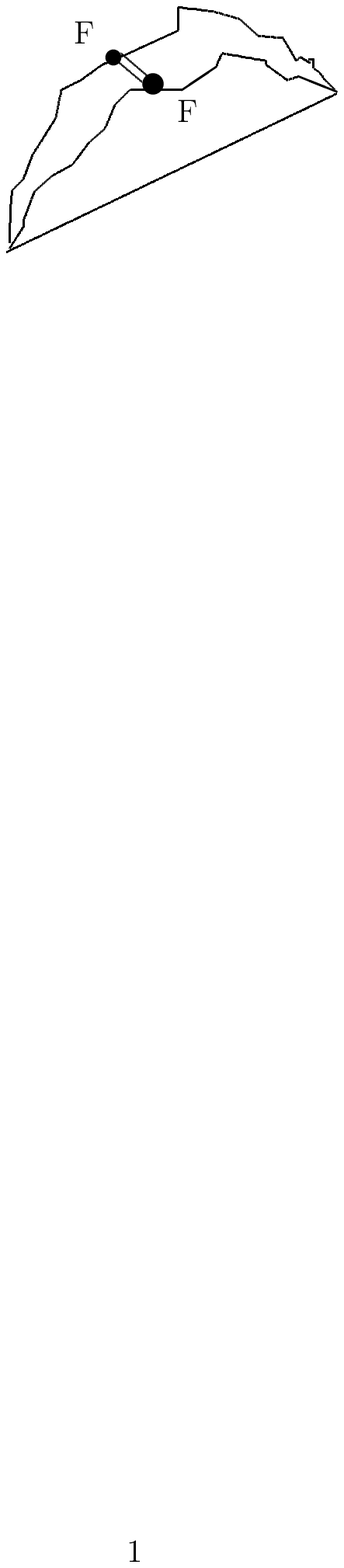}
\end{center}
\centerline{Fig. 5}
\end{figure}
\begin{figure}
\begin{center}
 \includegraphics[width=45mm,keepaspectratio=true]{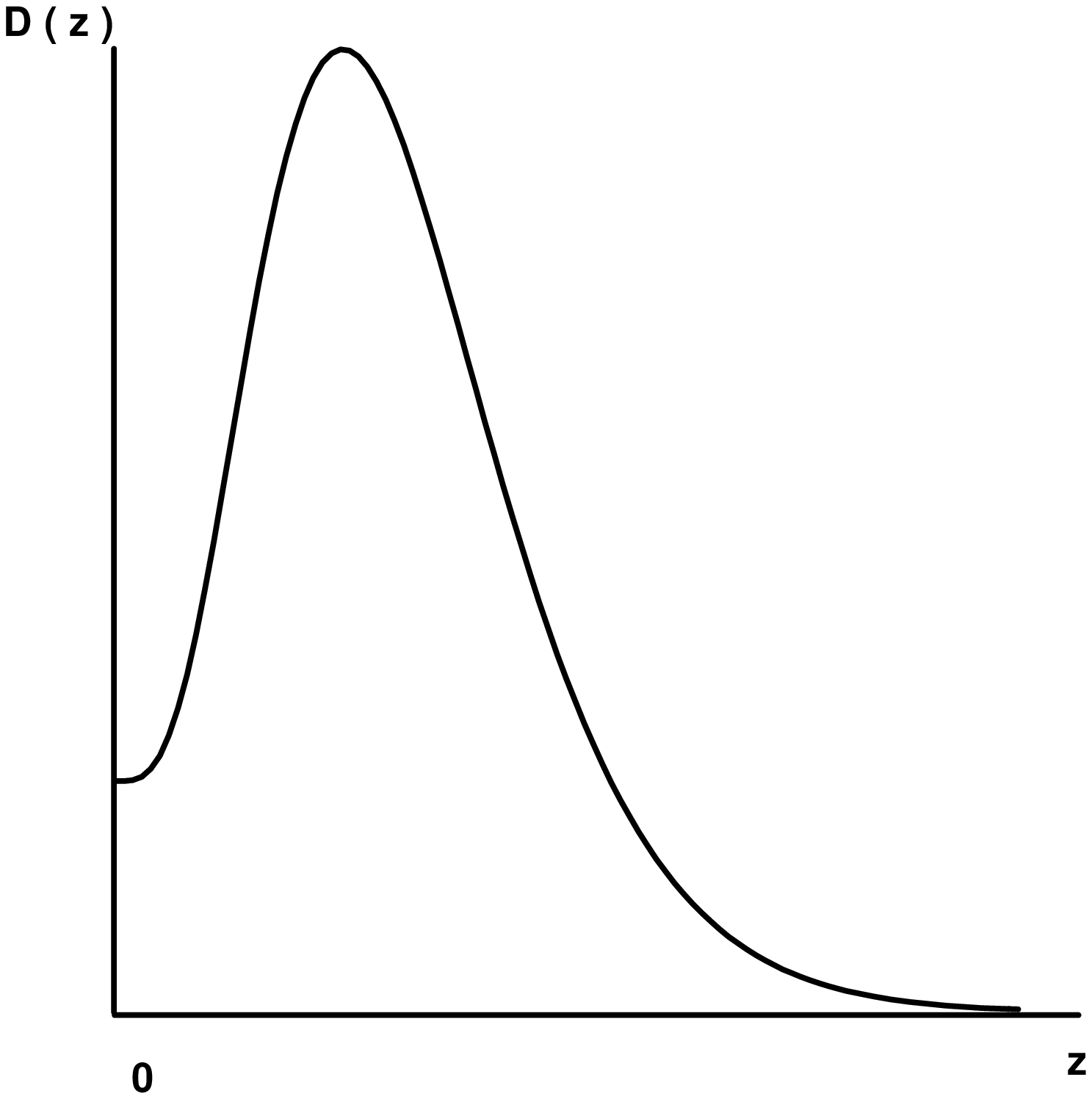}
\end{center}
\centerline{Fig. 6}
\end{figure}
\end{document}